\documentclass[twocolumn]{article}%
\usepackage{amsmath}
\usepackage{graphicx}%
\usepackage{amsfonts}%
\usepackage{amssymb}
\newtheorem{theorem}{Theorem}

\newtheorem{definition}[theorem]{Definition}

\newtheorem{lemma}[theorem]{Lemma}

\newtheorem{proposition}[theorem]{Proposition}

\newenvironment{proof}[1][Proof]{\noindent\textbf{#1.} }{\ \rule{0.5em}{0.5em}}
\begin{document}

\title{Quantum Certificate Complexity}
\author{Scott Aaronson\thanks{Email: \texttt{aaronson@cs.berkeley.edu}. \ Supported by
an NSF Graduate Fellowship and by the Defense Advanced Research Projects
Agency (DARPA) and Air Force Laboratory, Air Force Materiel Command, USAF,
under agreement number F30602-01-2-0524.}\\Computer Science Department\\University of California, Berkeley}
\date{}
\maketitle

\begin{abstract}
Given a Boolean function $f$, we study two natural generalizations of
the certificate complexity $C\left(  f\right)  $: the randomized certificate
complexity $RC\left(  f\right)  $\ and the quantum certificate complexity
$QC(f)$. \ Using Ambainis' adversary method, we exactly characterize
$QC\left(  f\right)  $ as the square root of $RC(f)$. \ We then use this
result to prove the new relation $R_{0}\left(  f\right)  =O\left(
Q_{2}\left(  f\right)  ^{2}Q_{0}\left(  f\right)  \log n\right)$ for total $f$, where
$R_{0}$, $Q_{2}$, and $Q_{0}$ are zero-error randomized, bounded-error
quantum, and zero-error quantum query complexities respectively. \ Finally we
give asymptotic gaps between the measures, including a total $f$ for which
$C(f)$ is superquadratic in $QC\left(  f\right)  $, and a symmetric partial
$f$ for which $QC\left(  f\right)  =O\left(  1\right)  $\ yet $Q_{2}\left(
f\right)  =\Omega\left(  n/\log n\right)  $.

\end{abstract}

\section{Background}

Most of what is known about the power of quantum computing can be cast in the
query or decision-tree model
\cite{aaronson,ambainis0,ambainis,bbbv,bbcmw,bcwz,bw,grover,shi,dewolf,dewolf2}%
. \ Here one counts only the number of queries to the input, not the number of
computational steps. \ The appeal of this model lies in its extreme
simplicity---in contrast to (say) the Turing machine model, one feels the
query model ought to be `completely understandable.' \ In spite of this, open
problems abound.

Let $f:Dom\left(  f\right)  \rightarrow\left\{  0,1\right\}  $\ be a Boolean
function with $Dom\left(  f\right)  \subseteq\left\{  0,1\right\}  ^{n}$, that
takes input $Y=y_{1}\ldots y_{n}$. \ Then the deterministic query complexity
$D\left(  f\right)  $\ is the minimum number of queries to the $y_{i}$'s
needed to evaluate $f$, if $Y$ is chosen adversarially and if queries can be
adaptive (that is, can depend on the outcomes of previous queries). \ Also,
the bounded-error randomized query complexity, $R_{2}\left(  f\right)  $, is
the minimum expected number of queries needed by a randomized algorithm that,
for each $Y\in Dom\left(  f\right)  $, outputs $f\left(  Y\right)  $ with
probability at least $2/3$. \ Here the `$2$' refers to two-sided error; if
instead we require $f\left(  Y\right)  $ to be output with probability $1$ for
every $Y$, we obtain $R_{0}\left(  f\right)  $, or zero-error randomized query complexity.

Analogously, $Q_{2}\left(  f\right)  $\ is the minimum number of queries
needed by a quantum algorithm that outputs $f\left(  Y\right)  $ with
probability at least $2/3$ for all $Y$. \ Also, for $k\in\left\{  0,1\right\}
$\ let $Q_{0}^{k}\left(  f\right)  $\ be the minimum number of queries needed
by a quantum algorithm that outputs $f\left(  Y\right)  $\ with probability
$1$ if $f\left(  Y\right)  =k$, and with probability at least $1/2$ if
$f\left(  Y\right)  \neq k$.\ \ Then let $Q_{0}\left(  f\right)  =\max\left\{
Q_{0}^{0}\left(  f\right)  ,Q_{0}^{1}\left(  f\right)  \right\}  $. \ If we
require a single algorithm that succeeds with probability $1$ for all $Y$, we
obtain $Q_{E}\left(  f\right)  $, or exact quantum query complexity. \ See
\cite{bw}\ for detailed definitions and a survey of these measures.

It is immediate that $Q_{2}\left(  f\right)  \leq R_{2}\left(  f\right)  \leq
R_{0}\left(  f\right)  \leq D\left(  f\right)  \leq n$, that $Q_{0}\left(
f\right)  \leq R_{0}\left(  f\right)  $, and that $Q_{E}\left(  f\right)  \leq
D\left(  f\right)  $. \ If $f$ is partial (i.e. $Dom\left(  f\right)
\neq\left\{  0,1\right\}  ^{n}$), then $Q_{2}\left(  f\right)  $ can be
superpolynomially smaller than $R_{2}\left(  f\right)  $; this is what makes
Shor's period-finding algorithm \cite{shor} possible. \ For total $f$, by
contrast, the largest known gap even between $D\left(  f\right)  $\ and
$Q_{2}\left(  f\right)  $\ is quadratic, and is achieved by the $OR$ function
on $n$ bits: $D\left(  OR\right)  =n$\ (indeed $R_{2}\left(  OR\right)
=\Omega\left(  n\right)  $), whereas $Q_{2}\left(  OR\right)  =\Theta\left(
\sqrt{n}\right)  $\ because of Grover's search algorithm \cite{grover}.
\ Furthermore, for total $f$, Beals et al. \cite{bbcmw}\ showed that $D\left(
f\right)  =O\left(  Q_{2}\left(  f\right)  ^{6}\right)  $, while de Wolf
\cite{dewolf2}\ showed that $D\left(  f\right)  =O\left(  Q_{2}\left(
f\right)  ^{2}Q_{0}\left(  f\right)  ^{2}\right)  $.

The result of Beals et al. \cite{bbcmw} relies on two intermediate complexity
measures, the \textit{certificate complexity} $C\left(  f\right)  $\ and
\textit{block sensitivity} $bs\left(  f\right)  $, which we now define.

\begin{definition}
A certificate for an input $X$ is a set $S\subseteq\left\{  1,\ldots
,n\right\}  $\ such that for all $Y\in Dom\left(  f\right)  $, if $y_{i}%
=x_{i}$\ for all $i\in S$\ then $f\left(  Y\right)  =f\left(  X\right)  $.
\ Then $C^{X}\left(  f\right)  $\ is the minimum size of a certificate for
$X$, and $C\left(  f\right)  $\ is the maximum of $C^{X}\left(  f\right)
$\ over all $X$.
\end{definition}

\begin{definition}
A sensitive block on input $X$ is a set $B\subseteq\left\{  1,\ldots
,n\right\}  $\ such that $f\left(  X^{\left(  B\right)  }\right)  \neq
f\left(  X\right)  $, where $X^{\left(  B\right)  }$\ is obtained from $X$ by
flipping $x_{i}$\ for each $i\in B$. \ Then $bs^{X}\left(  f\right)  $\ is the
maximum number of disjoint sensitive blocks on $X$, and $bs\left(  f\right)
$\ is the maximum of $bs^{X}\left(  f\right)  $\ over all $X$.
\end{definition}

Clearly $bs\left(  f\right)  \leq C\left(  f\right)  \leq D\left(  f\right)
$. \ For total $f$, these measures are all polynomially related: Nisan
\cite{nisan}\ showed that $C\left(  f\right)  \leq bs\left(  f\right)  ^{2}$,
while Beals et al. \cite{bbcmw}\ showed that $D\left(  f\right)  \leq C\left(
f\right)  bs\left(  f\right)  $. \ Combining these results with $bs\left(
f\right)  =O\left(  Q_{2}\left(  f\right)  ^{2}\right)  $ (from the optimality
of Grover's algorithm), one obtains $D\left(  f\right)  =O\left(  Q_{2}\left(
f\right)  ^{6}\right)  $.

\section{Our Results}

We investigate $RC\left(  f\right)  $ and $QC\left(  f\right)  $, the
bounded-error randomized and quantum generalizations of the certificate
complexity $C\left(  f\right)  $ (see Table 1). \ Our motivation is that, just
as $C\left(  f\right)  $\ was used to show a polynomial relation between
$D\left(  f\right)  $\ and $Q_{2}\left(  f\right)  $, so $RC\left(  f\right)
$ and $QC\left(  f\right)  $\ can lead to new relations among fundamental
query complexity measures.%
\[%
\begin{tabular}
[c]{l|ccc}%
\textbf{Table 1} &  &  & \\\hline
Query complexity & $D\left(  f\right)  $ & $R_{2}\left(  f\right)  $ &
$Q_{2}\left(  f\right)  $\\
Certificate complexity & $C\left(  f\right)  $ & $RC\left(  f\right)  $ &
$QC\left(  f\right)  $%
\end{tabular}
\ \ \
\]

What the certificate complexity $C\left(  f\right)  $\ measures is the number
of \textit{queries} used to verify a certificate, not the number of
\textit{bits} used to communicate it. \ Thus, if we want to generalize
$C\left(  f\right)  $, we should assume the latter is unbounded. \ A
consequence is that without loss of generality, a certificate is just a
claimed value $X$\ for the input $Y$\footnote{Throughout this paper, we use
$Y$ to denote the `actual' input being queried, and $X$ to denote the
`claimed' input (whose randomized certificate complexity, block sensitivity,
and so on we want to study).}---since any additional information that a prover
might provide, the verifier can compute for itself. \ The verifier's job is to
check that $f\left(  Y\right)  =f\left(  X\right)  $. \ With this in mind we
define $RC\left(  f\right)  $ as follows.

\begin{definition}
A randomized verifier for input $X$ is a randomized algorithm that, on input
$Y\in Dom\left(  f\right)  $, (i) accepts with probability $1$ if $Y=X$, and
(ii) rejects with probability at least $1/2$ if $f\left(  Y\right)  \neq
f\left(  X\right)  $. \ (If $Y\neq X$ but $f\left(  Y\right)  =f\left(
X\right)  $,\ the acceptance probability can be arbitrary.) $\ $Then
$RC^{X}\left(  f\right)  $\ is the minimum expected number of queries used by
a randomized verifier for $X$, and $RC\left(  f\right)  $\ is the maximum of
$RC^{X}\left(  f\right)  $\ over all $X$.
\end{definition}

We define $QC\left(  f\right)  $\ analogously, with quantum instead of
randomized algorithms. \ The following justifies the definition (the
$RC\left(  f\right)  $ part was originally shown by Raz et al. \cite{rtvv}).

\begin{proposition}
\label{onesided}Making the error probability two-sided rather than one-sided
changes $RC\left(  f\right)  $\ and $QC\left(  f\right)  $\ by at most a
constant factor.
\end{proposition}

\begin{proof}
For $RC\left(  f\right)  $, let $r_{V}^{Y}$\ be the event that verifier $V$
rejects on input $Y$, and let $d_{V}^{Y}$\ be the event that $V$ encounters a
disagreement with $X$ on $Y$. \ We may assume $\Pr\left[  r_{V}^{Y}%
\,|\,d_{V}^{Y}\right]  =1$. \ Suppose that $Y=X$ and $f\left(  Y\right)  \neq
f\left(  X\right)  $\ both occur with probability $1/2$, and that $\Pr\left[
r_{V}^{Y}\right]  \leq\varepsilon_{0}$ in the former case and $\Pr\left[
r_{V}^{Y}\right]  \geq1-\varepsilon_{1}$\ in the latter. \ Then%
\begin{align*}
\Pr\left[  \urcorner d_{V}^{Y}\,|\,r_{V}^{Y}\right]   &  =\frac{\Pr\left[
r_{V}^{Y}\,|\,\urcorner d_{V}^{Y}\right]  \Pr\left[  \urcorner d_{V}%
^{Y}\right]  }{\Pr\left[  r_{V}^{Y}\right]  }\\
&  \leq\frac{2\varepsilon_{0}}{1-\varepsilon_{1}}\text{.}%
\end{align*}
Now let $V^{\ast}$\ be identical to $V$ except that, whenever $V$ rejects
despite having found no disagreement with $X$, $V^{\ast}$\ accepts. \ Clearly
$\Pr\left[  r_{V^{\ast}}^{X}\right]  =0$. \ Also, in the case $f\left(
Y\right)  \neq f\left(  X\right)  $,%
\begin{align*}
\Pr\left[  r_{V^{\ast}}^{Y}\right]   &  =\Pr\left[  d_{V}^{Y}\right] \\
&  \geq\Pr\left[  r_{V}^{Y}\right]  \Pr\left[  d_{V}^{Y}\,|\,r_{V}^{Y}\right]
\\
&  \geq\left(  1-\varepsilon_{1}\right)  \left(  1-\frac{2\varepsilon_{0}%
}{1-\varepsilon_{1}}\right)  \text{.}%
\end{align*}

For $QC\left(  f\right)  $, suppose the verifier's final state given input $Y$
is%
\[
\sum_{z}\alpha_{z}^{Y}\left|  z\right\rangle \left(  \beta_{z}^{Y}\left|
0\right\rangle +\gamma_{z}^{Y}\left|  1\right\rangle \right)
\]
where $\left|  0\right\rangle $\ is the reject state, $\left|  1\right\rangle
$\ is the accept state, and $\left|  \beta_{z}^{Y}\right|  ^{2}+\left|
\gamma_{z}^{Y}\right|  ^{2}=1$\ for all $z$. \ Suppose also that $A^{X}%
\geq1-\varepsilon_{0}$\ and that $A^{Y}\leq\varepsilon_{1}$\ whenever
$f\left(  Y\right)  \neq f\left(  X\right)  $, where $A^{Y}=\sum_{z}\left|
\alpha_{z}^{Y}\gamma_{z}^{Y}\right|  ^{2}$ is the probability of accepting.
\ Then the verifier can make $A^{X}=1$\ by performing the conditional rotation%
\[
\left(
\begin{array}
[c]{cc}%
\gamma_{z}^{X} & -\beta_{z}^{X}\\
\beta_{z}^{X} & \gamma_{z}^{X}%
\end{array}
\right)
\]
on the second register prior to measurement. \ In the case $f\left(  Y\right)
\neq f\left(  X\right)  $, this produces%
\begin{align*}
A^{Y}  &  =\sum_{z}\left|  \alpha_{z}^{Y}\right|  ^{2}\left|  \beta_{z}%
^{X}\beta_{z}^{Y}+\gamma_{z}^{X}\gamma_{z}^{Y}\right|  ^{2}\\
&  \leq2\sum_{z}\left|  \alpha_{z}^{Y}\right|  ^{2}\left(  \left|  \beta
_{z}^{X}\right|  ^{2}+\left|  \gamma_{z}^{Y}\right|  ^{2}\right) \\
&  \leq2\left(  \varepsilon_{0}+\varepsilon_{1}\right)  \text{.}%
\end{align*}
\end{proof}

It is immediate that $QC\left(  f\right)  \leq RC\left(  f\right)  \leq
C\left(  f\right)  $, that $QC\left(  f\right)  =O\left(  Q_{2}\left(
f\right)  \right)  $, and that $RC\left(  f\right)  =O\left(  R_{2}\left(
f\right)  \right)  $. \ We also have $RC\left(  f\right)  =\Omega\left(
bs\left(  f\right)  \right)  $, since a randomized verifier for $X$ must query
each sensitive block on $X$ with $1/2$ probability. \ This suggests viewing
$RC\left(  f\right)  $\ as an `alloy' of block sensitivity and certificate
complexity, an interpretation for which Section \ref{gap}\ gives some justification.

Our results are as follows. \ In Section \ref{charsec}\ we show that
$QC\left(  f\right)  =\Theta\left(  \sqrt{RC\left(  f\right)  }\right)  $ for
all $f$ (partial or total), precisely characterizing quantum certificate
complexity in terms of randomized certificate complexity. To do this, we first
give a nonadaptive characterization of $RC\left(  f\right)  $, and then apply
the adversary method of Ambainis \cite{ambainis}\ to lower-bound $QC\left(
f\right)  $\ in terms of this characterization. \ Then, in Section \ref{rrc},
we extend results on polynomials due to de Wolf \cite{dewolf2}\ and to Nisan
and Smolensky (as described by Buhrman and de Wolf \cite{bw}), to show that
$R_{0}\left(  f\right)  =O\left(  RC\left(  f\right)  nd{}eg\left(  f\right)
\log n\right)  $\ for all total $f$, where $nd{}eg\left(  f\right)  $\ is the
minimum degree of a polynomial $p$\ such that $p\left(  X\right)  \neq0$\ if
and only if $f\left(  X\right)  \neq0$. \ Combining the results of Sections
\ref{charsec}\ and \ref{rrc} leads to a new lower bound on quantum query
complexity: that $R_{0}\left(  f\right)  =O\left(  Q_{2}\left(  f\right)
^{2}Q_{0}\left(  f\right)  \log n\right)  $ for all total $f$. \ To our
knowledge, this is the first quantum lower bound to use both the adversary
method and the polynomial method at different points in the argument.

Finally, in Section \ref{gap}, we exhibit asymptotic gaps between $RC\left(
f\right)  $\ and other query complexity measures, including a total $f$ for
which $C\left(  f\right)  =\Theta\left(  QC\left(  f\right)  ^{2.205}\right)
$, and a symmetric partial $f$ for which $QC\left(  f\right)  =O\left(
1\right)  $\ yet $Q_{2}\left(  f\right)  =\Omega\left(  n/\log n\right)  $.
\ We conclude in Section \ref{open}\ with some open problems.

\section{Related Work}

Raz et al. \cite{rtvv}\ studied a query complexity measure they called
$ma\left(  f\right)  $, for Merlin-Arthur. \ In our notation, $ma\left(
f\right)  $\ equals the maximum of $RC^{X}\left(  f\right)  $\ over all $X$
with $f\left(  X\right)  =1$. \ Raz et al. observed that $ma\left(  f\right)
=ip\left(  f\right)  $, where $ip\left(  f\right)  $\ is the number of queries
needed given arbitrarily many rounds of interaction with a prover. \ They also
used error-correcting codes to construct a total $f$ for which $ma\left(
f\right)  =O\left(  1\right)  $\ but $C\left(  f\right)  =\Omega\left(
n\right)  $. \ This has similarities to our construction, in Section
\ref{sym}, of a symmetric partial $f$ for which $QC\left(  f\right)  =O\left(
1\right)  $\ but $Q_{2}\left(  f\right)  =\Omega\left(  n/\log n\right)  $.
\ Aside from that and from Proposition \ref{onesided}, Raz et al.'s results do
not overlap with ours.

Watrous \cite{watrous}\ has investigated a different notion of `quantum
certificate complexity'---whether certificates that are quantum states can be
superpolynomially smaller than any classical certificate. \ Also, de Wolf
\cite{dewolf}\ has investigated `nondeterministic quantum query complexity' in
the alternate sense of algorithms that accept with zero probability when
$f\left(  Y\right)  =0$, and with positive probability when $f\left(
Y\right)  =1$.

\section{\label{charsec}Characterization of Quantum Certificate Complexity}

We wish to show that $QC\left(  f\right)  =\Theta\left(  \sqrt{RC\left(
f\right)  }\right)  $, precisely characterizing quantum certificate complexity
in terms of randomized certificate complexity. \ The first step is to give a
simpler characterization of $RC\left(  f\right)  $.

\begin{lemma}
Call a randomized verifier for $X$ \textit{nonadaptive} if, on input $Y$, it
queries each $y_{i}$\ with independent probability $\lambda_{i}$, and rejects
if and only if it encounters a disagreement with $X$. \ (Thus, we identify
such a verifier with the vector $\left(  \lambda_{1},\ldots,\lambda
_{n}\right)  $.) \ Let $RC_{na}^{X}\left(  f\right)  $\ be the minimum of
$\lambda_{1}+\cdots+\lambda_{n}$\ over all nonadaptive verifiers for
$X$.\ \ Then $RC_{na}^{X}\left(  f\right)  =\Theta\left(  RC^{X}\left(
f\right)  \right)  $.
\end{lemma}

\begin{proof}
Clearly $RC_{na}^{X}\left(  f\right)  =\Omega\left(  RC^{X}\left(  f\right)
\right)  $. \ For the upper bound, we can assume that a randomized verifier
rejects immediately on finding a disagreement with $X$, and accepts if it
finds no disagreement. \ Let $\mathcal{Y}=\left\{  Y:f\left(  Y\right)  \neq
f\left(  X\right)  \right\}  $. \ Let $V$ be an optimal randomized verifier,
and let $p_{t}\left(  Y\right)  $\ be the probability that $V$, when given
input $Y\in\mathcal{Y}$, finds a disagreement with $X$ on the $t^{th}$\ query.
\ By Markov's inequality, $V$ must have found a disagreement with probability
at least $1/2$ after $T=\left\lceil 2RC^{X}\left(  f\right)  \right\rceil
$\ queries. \ So by the union bound%
\[
p_{1}\left(  Y\right)  +\cdots+p_{T}\left(  Y\right)  \geq1/2
\]
for each $Y\in\mathcal{Y}$. \ Suppose we choose $t\in\left\{  1,\ldots
,T\right\}  $ uniformly at random and simulate the $t^{th}$\ query, pretending
that queries $1,\ldots,t-1$\ have already been made and have returned
agreement with $X$. \ Then we must find a disagreement with probability at
least $1/2T$. \ By repeating this procedure $4T$\ times, we can boost the
probability to $1-e^{-2}$. \ For $i\in\left\{  1,\ldots,n\right\}  $, let
$\lambda_{i}$\ be the probability that $y_{i}$\ is queried at least once.
\ Then $\lambda_{1}+\cdots+\lambda_{n}\leq4T$, whereas for each $Y\in
\mathcal{Y}$,%
\[
\sum_{i:y_{i}\neq x_{i}}\lambda_{i}\geq1-e^{-2}.
\]
It follows that, if each $y_{i}$\ is queried with independent probability
$\lambda_{i}$, then the probability that at least one $y_{i}$ disagrees with
$X$ is at least%
\[
1-\prod_{i:y_{i}\neq x_{i}}\left(  1-\lambda_{i}\right)  \geq1-\left(
1-\frac{1-e^{-2}}{n}\right)  ^{n}>0.57.
\]
\end{proof}

To obtain a lower bound on $QC\left(  f\right)  $, we use the following simple
reformulation of the adversary method of Ambainis \cite{ambainis}.

\begin{theorem}
[Ambainis]\label{ambthm}Let $\beta$ be a function from $Dom\left(  f\right)
$\ to nonnegative reals, and let $R:Dom\left(  f\right)  ^{2}\rightarrow
\left\{  0,1\right\}  $\ be a relation such that $R\left(  X,Y\right)
=R\left(  Y,X\right)  $\ for all $X,Y$\ and $R\left(  X,Y\right)
=0$\ whenever $f\left(  X\right)  =f\left(  Y\right)  $. \ Let $\delta
_{0},\delta_{1}\in\left(  0,1\right]  $\ be such that for every $X\in
Dom\left(  f\right)  $\ and $i\in\left\{  1,\ldots,n\right\}  $,%
\begin{align*}
\sum_{Y\,:\,R\left(  X,Y\right)  =1}\beta\left(  Y\right)   &  \geq1,\\
\sum_{Y\,:\,R\left(  X,Y\right)  =1,x_{i}\neq y_{i}}\beta\left(  Y\right)   &
\leq\delta_{f\left(  X\right)  }.
\end{align*}
Then $Q_{2}\left(  f\right)  =\Omega\left(  \sqrt{\frac{1}{\delta_{0}%
\delta_{1}}}\right)  $.
\end{theorem}

We now prove the main result of the section.

\begin{theorem}
\label{sqrt}For all $f$ (partial or total) and all $X$, $QC^{X}\left(
f\right)  =\Theta\left(  \sqrt{RC^{X}\left(  f\right)  }\right)  $.
\end{theorem}

\begin{proof}
Let $\left(  \lambda_{1},\ldots,\lambda_{n}\right)  $ be an optimal
nonadaptive randomized verifier for $X$, and let%
\[
S=\lambda_{1}+\cdots+\lambda_{n}.
\]
First, $QC^{X}\left(  f\right)  =O\left(  \sqrt{S}\right)  $. \ We can run a
``weighted Grover search,'' in which the proportion of basis states querying
index $i$ is within a constant factor of $\lambda_{i}/S$. \ (It suffices to
use $n^{2}$\ basis states.) \ Let $\mathcal{Y}=\left\{  Y:f\left(  Y\right)
\neq f\left(  X\right)  \right\}  $;\ then for any $Y\in\mathcal{Y}$,
$O\left(  \sqrt{S}\right)  $\ iterations suffice to find a disagreement with
$X$ with probability $\Omega\left(  1\right)  $.

Second, $QC^{X}\left(  f\right)  =\Omega\left(  \sqrt{S}\right)  $. \ Consider
a matrix game in which Alice chooses an index $i$\ to query and Bob chooses
$Y\in\mathcal{Y}$; Alice wins if and only if $y_{i}\neq x_{i}$. \ If both
players are rational, then Alice wins with probability $O\left(  1/S\right)
$, since otherwise Alice's strategy would yield a verifier $\left(
\lambda_{1}^{\prime},\ldots,\lambda_{n}^{\prime}\right)  $\ with%
\[
\lambda_{1}^{\prime}+\cdots+\lambda_{n}^{\prime}=o\left(  S\right)  .
\]
Hence by the minimax theorem, there exists a distribution $\mu$\ over
$\mathcal{Y}$\ such that for every $i$,%
\[
\Pr_{Y\in\mu}\left[  y_{i}\neq x_{i}\right]  =O\left(  1/S\right)  .
\]
Let $\beta\left(  X\right)  =1$\ and let $\beta\left(  Y\right)  =\mu\left(
Y\right)  $\ for each $Y\in\mathcal{Y}$. \ Also, let $R\left(  Y,Z\right)
=1$\ if and only if $Z=X$ for each $Y\in\mathcal{Y}$ and $Z\notin\mathcal{Y}$.
\ Then we can take $\delta_{f\left(  Y\right)  }=1$\ and $\delta_{f\left(
X\right)  }=O\left(  1/S\right)  $ in Theorem \ref{ambthm}. \ So the quantum
query complexity of distinguishing $X$ from an arbitrary $Y\in\mathcal{Y}$\ is
$\Omega\left(  \sqrt{S}\right)  $.
\end{proof}

\section{\label{rrc}Quantum Lower Bound for Total Functions}

Our goal is to show that%
\[
R_{0}\left(  f\right)  =O\left(  Q_{2}\left(  f\right)  ^{2}Q_{0}\left(
f\right)  \log n\right)  .
\]
Say that a real multilinear polynomial $p\left(  x_{1},\ldots,x_{n}\right)
$\ nondeterministically represents $f$ if for all $X\in\left\{  0,1\right\}
^{n}$, $p\left(  X\right)  \neq0$\ if and only if $f\left(  X\right)  \neq
0$.\ \ Let $nd{}eg\left(  f\right)  $ be the minimum degree of a
nondeterministic polynomial for $f$. \ Also, given such a polynomial $p$, say
that a monomial $M_{1}\in p$\ is \textit{covered} by $M_{2}\in p$\ if $M_{2}%
$\ contains every variable in $M_{1}$.\ \ We call $M$ a \textit{maxonomial} if
it is not covered by any other monomial of $p$. \ The following is a simple
generalization of a lemma attributed in \cite{bw} to Nisan and Smolensky.

\begin{lemma}
[Nisan-Smolensky]\label{maxo}Let $p$ nondeterministically represent $f$.
\ Then for every maxonomial $M$ of $p$ and $X\in f^{-1}\left(  0\right)  $,
there is a set $B$ of variables in $M$ such that $f\left(  X^{\left(
B\right)  }\right)  \neq f\left(  X\right)  $, where $X^{\left(  B\right)  }%
$\ is obtained from $X$\ by flipping the variables in $B$.
\end{lemma}

\begin{proof}
Obtain a restricted function $g$ from $f$, and a restricted polynomial $q$
from $p$, by setting each variable outside of $M$ to $x_{i}$. \ Then $g$
cannot be constant, since its representing polynomial $q$ contains $M$ as a
monomial. \ Thus there is a subset $B$ of variables in $M$ such that $g\left(
X^{\left(  B\right)  }\right)  =1$, and hence $f\left(  X^{\left(  B\right)
}\right)  =1$.
\end{proof}

Using Lemma \ref{maxo}, de Wolf \cite{dewolf2}\ showed that $D\left(
f\right)  \leq C\left(  f\right)  nd{}eg\left(  f\right)  $ for all total $f$
(slightly improving the result $D\left(  f\right)  \leq C\left(  f\right)
\deg\left(  f\right)  $\ due to Buhrman and de Wolf \cite{bw}). \ In Theorem
\ref{degthm}, we will give an analog of this result for \textit{randomized}
query and certificate complexities. \ However, we first need a probabilistic lemma.

\begin{lemma}
\label{undom}Suppose we repeatedly apply the following procedure: first
identify the set $B$\ of maxonomials of $p$, then `shrink' each $M\in B$ with
(not necessarily independent) probability at least $1/2$. \ Shrinking $M$
means replacing it by an arbitrary monomial of degree $\deg\left(  M\right)
-1$. \ Then with high probability $p$ is a constant polynomial after $O\left(
\deg\left(  p\right)  \log n\right)  $\ iterations.
\end{lemma}

\begin{proof}
For any set $A$ of monomials, consider the weighting function%
\[
\omega\left(  A\right)  =\sum_{M\in A}\deg\left(  M\right)  !
\]
Let $S$ be the set of monomials of $p$. \ Initially%
\[
\omega\left(  S\right)  \leq n^{\deg\left(  p\right)  }\deg\left(  p\right)
!
\]
and we are done when $\omega\left(  S\right)  =0$. \ We claim that at every
iteration, $\omega\left(  B\right)  \geq\frac{1}{e}\omega\left(  S\right)  $.
\ For every $M^{\ast}\in S\setminus B$\ is covered by some $M\in B$, but a
given $M\in B$\ can cover at most $\tbinom{\deg\left(  M\right)  }{l}%
$\ distinct $M^{\ast}$ with $\deg\left(  M^{\ast}\right)  =l$. \ Hence%
\begin{align*}
\omega\left(  S\setminus B\right)   &  \leq\sum_{M\in B}\sum_{l=0}%
^{\deg\left(  M\right)  -1}\tbinom{\deg\left(  M\right)  }{l}l!\\
&  \leq\sum_{M\in B}\deg\left(  M\right)  !\left(  \frac{1}{1!}+\frac{1}%
{2!}+\cdots\right) \\
&  \leq\left(  e-1\right)  \omega\left(  B\right)  .
\end{align*}
At every iteration, the contribution of each $M\in B$\ to $\omega\left(
A\right)  $\ has at least $1/2$ probability of shrinking from $\deg\left(
M\right)  !$\ to $\left(  \deg\left(  M\right)  -1\right)  !$\ (or to $0$ if
$\deg\left(  M\right)  =1$). \ Hence $\omega\left(  S\right)  $\ decreases by
an expected amount at least $\frac{1}{4e}\omega\left(  S\right)  $. \ Thus
after%
\[
\log_{4e/\left(  4e-1\right)  }\left(  2n^{\deg\left(  p\right)  }\deg\left(
p\right)  !\right)  =O\left(  \deg\left(  p\right)  \log n\right)
\]
iterations, the expectation of $\omega\left(  S\right)  $\ is less than $1/2$,
so $S$ is empty with probability at least $1/2$.
\end{proof}

\begin{theorem}
\label{degthm}For total $f$,%
\[
R_{0}\left(  f\right)  =O\left(  RC\left(  f\right)  nd{}eg\left(  f\right)
\log n\right)  .
\]
\end{theorem}

\begin{proof}
Choose an $X$ with $f\left(  X\right)  =0$, and\ let $\left(  \lambda
_{1},\ldots,\lambda_{n}\right)  $\ be a nonadaptive randomized verifier for
$X$. \ Form $I\subseteq\left\{  1,\ldots,n\right\}  $\ by placing each $i$ in
$I$\ with independent probability $\lambda_{i}$. \ Then for any $Z\in\left\{
0,1\right\}  ^{n}$, let $Z^{\left[  I\right]  }$\ be obtained from $Z$ by
setting $z_{i}$\ to $x_{i}$\ for each $i\in I$. \ We have $\Pr_{I}\left[
f\left(  Z^{\left[  I\right]  }\right)  =0\right]  \geq1/2$. \ But by Lemma
\ref{maxo}, for every maxonomial $M$ of $f$, there exists a $Z$ that disagrees
with $X$ only on variables occurring in $M$, such that $f\left(  Z\right)
=1$. \ It follows that for every $M$, $I$ contains the index of a variable in
$M$ with probability at least $1/2$.

Given input $Y$, the randomized algorithm is as follows. \ First query the
indices in $I$, and let $f_{1}$\ be the restriction of $f$ induced by this.
\ Then repeat the above procedure on $f_{1}$---that is, choose an $X_{1}%
$\ with $f_{1}\left(  X_{1}\right)  =0$ (assuming one exists), and then query
a set $I_{1}$ drawn using a nonadaptive randomized verifier for $X_{1}$.
\ Continue in this manner until $f$ is restricted to a constant function
$f_{T}$. \ At this point, if $f_{T}$ is identically $0$ then we know $f\left(
Y\right)  =0$; otherwise we know $f\left(  Y\right)  =1$.

Each iteration of the algorithm uses an expected number of queries at most
$RC\left(  f\right)  $, since $RC\left(  g\right)  \leq RC\left(  f\right)
$\ for every restriction $g$ of $f$. \ Furthermore, since an iteration shrinks
each maxonomial with probability at least $1/2$, Lemma \ref{undom}\ implies
that with $\Omega\left(  1\right)  $\ probability, $f_{T}$\ is constant after
$T=O\left(  nd{}eg\left(  f\right)  \log n\right)  $\ iterations.
\end{proof}

Buhrman et al. \cite{bbcmw}\ showed that $nd{}eg\left(  f\right)  \leq
2Q_{0}\left(  f\right)  $. \ Combining this with Theorems \ref{sqrt}\ and
\ref{degthm}, we obtain a new relation between classical and quantum query complexity.

\begin{theorem}
\label{rqqe}For total $f$,%
\[
R_{0}\left(  f\right)  =O\left(  Q_{2}\left(  f\right)  ^{2}Q_{0}\left(
f\right)  \log n\right)  .
\]
\end{theorem}

The best previous relation of this kind was $R_{0}\left(  f\right)  =O\left(
Q_{2}\left(  f\right)  ^{2}Q_{0}\left(  f\right)  ^{2}\right)  $, due to de
Wolf \cite{dewolf2}.

\section{\label{gap}Asymptotic Gaps}

Having related $RC\left(  f\right)  $\ and $QC\left(  f\right)  $\ to other
query complexity measures in Section \ref{rrc}, in what follows we seek the
largest possible asymptotic gaps among the measures. \ In particular, Section
\ref{total}\ gives a total $f$ for which $RC\left(  f\right)  =\Theta\left(
C\left(  f\right)  ^{0.907}\right)  $ and hence $C\left(  f\right)
=\Theta\left(  QC\left(  f\right)  ^{2.205}\right)  $, as well as a total $f$
for which $bs\left(  f\right)  =\Theta\left(  RC\left(  f\right)
^{0.922}\right)  $. \ Although these gaps are the largest of which we know,
Section \ref{local}\ shows that no `local' technique can improve the relations
$C\left(  f\right)  =O\left(  RC\left(  f\right)  ^{2}\right)  $\ and
$RC\left(  f\right)  =O\left(  bs\left(  f\right)  ^{2}\right)  $. \ Finally,
Section \ref{sym}\ uses combinatorial designs to construct a symmetric partial
$f$ for which $RC\left(  f\right)  $\ and $QC\left(  f\right)  $ are $O\left(
1\right)  $,\ yet $Q_{2}\left(  f\right)  =\Omega\left(  n/\log n\right)  $.

\subsection{\label{total}Certificate Complexity, Randomized Certificate
Complexity, and Block Sensitivity}

Wegener and Z\'{a}dori \cite{wz}\ exhibited total Boolean functions with
asymptotic gaps between $C\left(  f\right)  $\ and $bs\left(  f\right)  $.
\ In similar fashion, we give a function family $\left\{  g_{t}\right\}  $
with an asymptotic gap between $C\left(  g_{t}\right)  $\ and $RC\left(
g_{t}\right)  $. \ Let $g_{1}\left(  x_{1},\ldots,x_{29}\right)  $\ equal $1$
if and only if the Hamming weight of its input is $13$, $14$, $15$, or $16$.
\ (The parameter $29$ was found via computer search to produce a maximal
separation.) \ Then for $t>1$, let%
\[
g_{t}\left(  x_{1},\ldots,x_{29^{t}}\right)  =g_{0}\left[  g_{t-1}\left(
X_{1}\right)  ,\ldots,g_{t-1}\left(  X_{29}\right)  \right]
\]
where $X_{1}$\ is the first $29^{t-1}$\ input bits, $X_{2}$\ is the second
$29^{t-1}$, and so on. \ For $k\in\left\{  0,1\right\}  $, let%
\begin{align*}
bs^{k}\left(  f\right)   &  =\max_{f\left(  X\right)  =k}bs^{X}\left(
f\right)  ,\\
C^{k}\left(  f\right)   &  =\max_{f\left(  X\right)  =k}C^{X}\left(  f\right)
.
\end{align*}
\ Then since $bs^{0}\left(  g_{1}\right)  =bs^{1}\left(  g_{1}\right)  =17$,
we have $bs\left(  g_{t}\right)  =17^{t}$. \ On the other hand, $C^{0}\left(
g_{1}\right)  =17$ but $C^{1}\left(  g_{1}\right)  =26$, so%
\begin{align*}
C^{1}\left(  g_{t}\right)   &  =13C^{1}\left(  g_{t-1}\right)  +13C^{0}\left(
g_{t-1}\right)  ,\\
C^{0}\left(  g_{t}\right)   &  =17\max\left\{  C^{1}\left(  g_{t-1}\right)
,C^{0}\left(  g_{t-1}\right)  \right\}  .
\end{align*}
Solving this recurrence yields $C\left(  g_{t}\right)  =\Theta\left(
22.725^{t}\right)  $. \ We can now show a gap between $C$\ and $RC$.

\begin{proposition}
\label{gapprop}$RC\left(  g_{t}\right)  =\Theta\left(  C\left(  g_{t}\right)
^{0.907}\right)  $.
\end{proposition}

\begin{proof}
Since $bs\left(  g_{t}\right)  =\Omega\left(  C\left(  g_{t}\right)
^{0.907}\right)  $, it suffices to show that $RC\left(  g_{t}\right)
=O\left(  bs\left(  g_{t}\right)  \right)  $. \ The randomized verifier
$V$\ chooses an input variable to query as follows. \ Let $X$\ be the claimed
input, and let $K=\sum_{i=1}^{29}g_{t-1}\left(  X_{i}\right)  $. \ Let
$I_{0}=\left\{  i:g_{t-1}\left(  X_{i}\right)  =0\right\}  $\ and
$I_{1}=\left\{  i:g_{t-1}\left(  X_{i}\right)  =1\right\}  $. \ With
probability $p_{K}$, $V$ chooses an $i\in I_{1}$\ uniformly at random;
otherwise $A$ chooses an $i\in I_{0}$\ uniformly at random. \ Here $p_{K}$\ is
as follows.%
\[%
\begin{tabular}
[c]{c|cccccc}%
$K$ & $\left[  0,12\right]  $ & $13$ & $14$ & $15$ & $16$ & $\left[
17,29\right]  $\\\hline
$p_{K}$ & $0$ & $\frac{13}{17}$ & $\frac{7}{12}$ & $\frac{5}{12}$ &
$\frac{4}{17}$ & $1$%
\end{tabular}
\
\]
Once $i$ is chosen, $V$ repeats the procedure for $X_{i}$, and\ continues
recursively in this manner until reaching a variable $y_{j}$ to query. \ One
can check that if $g_{t}\left(  X\right)  \neq g_{t}\left(  Y\right)  $, then
$g_{t-1}\left(  X_{i}\right)  \neq g_{t-1}\left(  Y_{i}\right)  $\ with
probability at least $1/17$. \ Hence $x_{j}\neq y_{j}$\ with probability at
least $1/17^{t}$, and $RC\left(  g_{t}\right)  =O\left(  17^{t}\right)  $.
\end{proof}

By Theorem \ref{sqrt}, it follows that $C\left(  g_{t}\right)  =\Theta\left(
QC\left(  g_{t}\right)  ^{2.205}\right)  $. \ This offers a surprising
contrast with the query complexity setting, where the best known gap between
the deterministic and quantum measures is quadratic ($D\left(  f\right)
=\Theta\left(  Q_{2}\left(  f\right)  ^{2}\right)  $).

The family $\left\{  g_{t}\right\}  $\ happens \textit{not} to yield an
asymptotic gap between $bs\left(  f\right)  $\ and $RC\left(  f\right)  $.
\ The reason is that any input to $g_{0}$\ can be covered perfectly by
sensitive blocks of minimum size, with no variables left over. In general,
though, we can have $bs\left(  f\right)  =o\left(  RC\left(  f\right)
\right)  $. \ As reported by Bublitz et al. \cite{bsvw}, M. Paterson found a
total Boolean function $h_{1}\left(  x_{1},\ldots,x_{6}\right)  $\ such that
$C^{X}\left(  h_{1}\right)  =5$\ and $bs^{X}\left(  h_{1}\right)  =4$\ for all
$X$. \ Composing $h_{1}$\ recursively yields $bs\left(  h_{t}\right)
=\Theta\left(  C\left(  h_{t}\right)  ^{0.861}\right)  $\ and $bs\left(
h_{t}\right)  =\Theta\left(  RC\left(  h_{t}\right)  ^{0.922}\right)  $, both
of which are the largest such gaps of which we know.

\subsection{\label{local}Local Separations}

It is a longstanding open question whether the relation $C\left(  f\right)
\leq bs\left(  f\right)  ^{2}$\ due to Nisan \cite{nisan}\ is tight. \ As a
first step, one can ask whether the relations $C\left(  f\right)  =O\left(
RC\left(  f\right)  ^{2}\right)  $\ and $RC\left(  f\right)  =O\left(
bs\left(  f\right)  ^{2}\right)  $\ are tight. \ In this section we introduce
a notion of \textit{local proof} in query complexity,\ and then show there is
no local proof that $C\left(  f\right)  =o\left(  RC\left(  f\right)
^{2}\right)  $\ or that $RC\left(  f\right)  =o\left(  bs\left(  f\right)
^{2}\right)  $. \ This implies that proving either result would require
techniques unlike those that are currently known. \ Our inspiration comes from
computational complexity, where researchers first formalized known methods of
proof, including \textit{relativizable proofs} \cite{bgs}\ and \textit{natural
proofs} \cite{rr}, and then argued that these methods were not powerful enough
to resolve the field's outstanding problems.

Let $G\left(  f\right)  $\ and $H\left(  f\right)  $\ be query complexity
measures obtained by maximizing over all inputs---that is, $G\left(  f\right)
=\max_{X\in Dom\left(  f\right)  }G^{X}\left(  f\right)  $\ and $H\left(
f\right)  =\max_{X\in Dom\left(  f\right)  }H^{X}\left(  f\right)  $. \ Call
$B\subseteq\left\{  1,\ldots,n\right\}  $\ a \textit{minimal block} on $X$ if
$B$ is sensitive on $X$ (meaning $f\left(  X^{\left(  B\right)  }\right)  \neq
f\left(  X\right)  $), and no sub-block $B^{\prime}\subset B$\ is sensitive on
$X$. \ Also, let $X$'s \textit{neighborhood} $\mathcal{N}\left(  X\right)
$\ consist of $X$ together with $X^{\left(  B\right)  }$\ for every minimal
block $B$ of $X$.\ \ Consider a proof that $G\left(  f\right)  =O\left(
t\left(  H\left(  f\right)  \right)  \right)  $ for some nondecreasing $t$.
\ We call the proof \textit{local} if it proceeds by showing that for every
$X\in Dom\left(  f\right)  $,%
\[
G^{X}\left(  f\right)  =O\left(  \max_{Y\in\mathcal{N}\left(  X\right)
}\left\{  t\left(  H^{Y}\left(  f\right)  \right)  \right\}  \right)  .
\]

As a canonical example, Nisan's proof \cite{nisan}\ that $C\left(  f\right)
\leq bs\left(  f\right)  ^{2}$\ is local. \ For each $X$, Nisan observes that
(i) a maximal set of disjoint minimal blocks is a certificate for $X$, (ii)
such a set can contain at most $bs^{X}\left(  f\right)  $\ blocks, and (iii)
each block can have size at most $\max_{Y\in\mathcal{N}\left(  X\right)
}bs^{Y}\left(  f\right)  $. \ Another example of a local proof is our proof in
Section \ref{charsec}\ that $RC\left(  f\right)  =O\left(  QC\left(  f\right)
^{2}\right)  $.

\begin{proposition}
\label{localprop}There is no local proof that $C\left(  f\right)  =o\left(
RC\left(  f\right)  ^{2}\right)  $\ or that $RC\left(  f\right)  =o\left(
bs\left(  f\right)  ^{2}\right)  $ for total $f$.
\end{proposition}

\begin{proof}
The first part is easy:\ let $f\left(  X\right)  =1$\ if $\left|  X\right|
\geq\sqrt{n}$\ (where $\left|  X\right|  $\ denotes the Hamming weight of
$X$), and $f\left(  X\right)  =0$ otherwise. \ Consider the all-zero input
$0^{n}$. \ We have $C^{0^{n}}\left(  f\right)  =n-\left\lceil \sqrt
{n}\right\rceil +1$, but $RC^{0^{n}}\left(  f\right)  =O\left(  \sqrt
{n}\right)  $, and indeed $RC^{Y}\left(  f\right)  =O\left(  \sqrt{n}\right)
$\ for all $Y\in\mathcal{N}\left(  0^{n}\right)  $.

For the second part, arrange the input variables in a lattice of size
$\sqrt{n}\times\sqrt{n}$. \ Take $m=\Theta\left(  n^{1/3}\right)  $, and let
$g\left(  X\right)  $\ be the monotone Boolean function that outputs $1$ if
and only if $X$ contains a $1$\textit{-square} of size $m\times m$. \ This is
a square of $1$'s that can wrap around the edges of the lattice; note that
only the variables along the sides must be set to $1$, not those in the
interior. \ An example input, with a $1$-square of size $3\times3$, is shown
below.%
\[%
\begin{array}
[c]{ccccc}%
0 & 0 & 0 & 0 & 0\\
0 & 0 & 0 & 0 & 0\\
1 & 0 & 0 & 1 & 1\\
1 & 0 & 0 & 1 & 0\\
1 & 0 & 0 & 1 & 1
\end{array}
\]
Clearly $bs^{0^{n}}\left(  g\right)  =\Theta\left(  n^{1/3}\right)  $, since
there can be at most $n/m^{2}$\ disjoint $1$-squares of size $m\times m$.
\ Also, $bs^{Y}\left(  g\right)  =\Theta\left(  n^{1/3}\right)  $\ for any
$Y$\ that is $0$ except for a single $1$-square. \ On the other hand, if we
choose uniformly at random among all such $Y$'s, then at any lattice site $i$,
$\Pr_{Y}\left[  y_{i}=1\right]  =\Theta\left(  n^{-2/3}\right)  $. \ Hence
$RC^{0^{n}}\left(  g\right)  =\Omega\left(  n^{2/3}\right)  $.
\end{proof}

\subsection{\label{sym}Symmetric Partial Functions}

If $f$ is partial, then $QC\left(  f\right)  $\ can be much smaller than
$Q_{2}\left(  f\right)  $. \ This is strikingly illustrated by the
\textit{collision problem}: let $Y=\left(  y_{1},\ldots,y_{n}\right)  $\ be a
sequence of integers in the range $\left\{  1,\ldots,n^{2}\right\}  $, each of
which\ can be retrieved by a single query. \ Let $Col\left(  Y\right)  =0$ if
$Y$\ is one-to-one (each $y_{i}$ is unique), and $Col\left(  Y\right)  =1$ if
$Y$ is two-to-one (each $y_{i}$\ appears exactly twice), under the promise
that one of these is the case. \ Then $RC\left(  Col\right)  =QC\left(
Col\right)  =O\left(  1\right)  $, since every one-to-one input differs from
every two-to-one input on at least $n/2$ of the $y_{i}$'s. \ On the other
hand, Aaronson \cite{aaronson}\ showed that $Q_{2}\left(  Col\right)
=\Omega\left(  n^{1/5}\right)  $, and Shi \cite{shi}\ improved this to
$\Omega\left(  n^{1/3}\right)  $,\ which is tight \cite{bht}.

From the example of the collision problem, it is tempting to conjecture
that\ (say) $Q_{2}\left(  f\right)  =O\left(  n^{1/3}\right)  $\ whenever
$QC\left(  f\right)  =O\left(  1\right)  $---that is, `if every $0$-input is
far from every $1$-input,\ then the quantum query complexity is sublinear.'
\ Here we disprove this conjecture, even for the special case of symmetric
functions such as $Col$. \ (For a finite set $\mathcal{H}$, we say that
$f:\mathcal{H}^{n}\rightarrow\left\{  0,1\right\}  $ is symmetric if
$y_{1}\ldots y_{n}\in Dom\left(  f\right)  $ implies $y_{\sigma\left(
1\right)  }\ldots y_{\sigma\left(  n\right)  }\in Dom\left(  f\right)  $ and
$f\left(  y_{1}\ldots x_{n}\right)  =f\left(  y_{\sigma\left(  1\right)
}\ldots y_{\sigma\left(  n\right)  }\right)  $ for every permutation $\sigma$.)

Our proof uses the following lemma, due to Nisan and Wigderson \cite{nw}.

\begin{lemma}
[Nisan-Wigderson]\label{nwlemma}For any $\gamma>1$, there exists a family of
sets%
\[
S_{1},\ldots,S_{m}\subseteq\left\{  1,\ldots,\left\lceil \gamma n\right\rceil
\right\}
\]
such that $m=\Omega\left(  2^{n/\gamma}\right)  $, $\left|  S_{i}\right|  =n$
for all $i$, and $\left|  S_{i}\cap S_{j}\right|  \leq n/\gamma$\ for all
$i\neq j$.
\end{lemma}

We will also need to adapt a lemma of Ambainis \cite{ambainis0}. \ For
$Z\in\left\{  0,1\right\}  ^{N}$, say that a multivariate polynomial $p\left(
Z\right)  $\ approximates $g\left(  Z\right)  $ if (i) $p\left(  Z\right)
\in\left[  0,1\right]  $\ for \textit{every} input $Z$ (not merely those in
$Dom\left(  f\right)  $), and (ii) $\left|  p\left(  Z\right)  -g\left(
Z\right)  \right|  \leq1/3$\ for every $Z\in Dom\left(  f\right)  $. \ Also,
let $\Delta\left(  N,d\right)  =\sum_{i=0}^{d}\tbinom{N}{i}$.

\begin{lemma}
[Ambainis]\label{amblemma}At most $2^{O\left(  \Delta\left(  N,d\right)
dN^{2}\right)  }$\ distinct Boolean functions (partial or total) can be
approximated by polynomials of degree $d$.
\end{lemma}

We can now prove the main result.

\begin{theorem}
\label{symthm}There exists a symmetric partial $f$ for which $QC\left(
f\right)  =O\left(  1\right)  $\ and $Q_{2}\left(  f\right)  =\Omega\left(
n/\log n\right)  $.
\end{theorem}

\begin{proof}
Let $f:\mathcal{H}^{n}\rightarrow\left\{  0,1\right\}  $\ where $\mathcal{H}%
=\left\{  1,\ldots,3n\right\}  $, and let $m=\Omega\left(  2^{n/3}\right)  $.
\ Let $S_{1},\ldots,S_{m}\subseteq\mathcal{H}$\ be as in Lemma \ref{nwlemma}.
\ We put $\left(  y_{1},\ldots,y_{n}\right)  $\ in $Dom\left(  f\right)  $\ if
and only if $\left\{  y_{1},\ldots,y_{n}\right\}  =S_{j}$\ for some $j$.
\ Clearly $QC\left(  f\right)  =O\left(  1\right)  $, since if $i\neq j$\ then
every permutation of $S_{i}$\ differs from every permutation of $S_{j}$\ on at
least $n/3$\ indices.

The number of symmetric $f$ with $Dom\left(  f\right)  $\ as above is
$2^{m}=2^{\Omega\left(  2^{n/3}\right)  }$. \ We can convert any such $f$ to a
Boolean function $g$ on $O\left(  n\log n\right)  $ variables. \ But Beals et
al. \cite{bbcmw}\ showed that, if $Q_{2}\left(  g\right)  =T$, then $g$ is
approximated by a polynomial of degree at most $2T$. \ So by Lemma
\ref{amblemma}, if $Q_{2}\left(  g\right)  \leq T$\ for every $g$ then%
\[
2T\cdot\Delta\left(  n\log n,2T\right)  \cdot\left(  n\log n\right)
^{2}=\Omega\left(  2^{n/3}\right)
\]
and we solve to obtain $T=\Omega\left(  n/\log n\right)  $.
\end{proof}

\section{\label{open}Open Problems}

Is $\widetilde{\deg}\left(  f\right)  =\Omega\left(  \sqrt{RC\left(  f\right)
}\right)  $, where $\widetilde{\deg}\left(  f\right)  $\ is the minimum degree
of a polynomial approximating $f$? \ In other words, can one lower-bound
$QC\left(  f\right)  $\ using the polynomial method of Beals et al.
\cite{bbcmw}, rather than the adversary method of Ambainis \cite{ambainis}?

Also, is $R_{2}\left(  f\right)  =O\left(  RC\left(  f\right)  ^{2}\right)  $?
\ If so we obtain the new relation $R_{2}\left(  f\right)  =O\left(
Q_{2}\left(  f\right)  ^{4}\right)  $.

\section{Acknowledgments}

I thank Ronald de Wolf for comments on the manuscript and for pointing out
that $Q_{E}\left(  f\right)  $\ can be replaced by $Q_{0}\left(  f\right)
$\ in Theorem \ref{rqqe}; and Umesh Vazirani and Ashwin Nayak for helpful discussions.

\end{document}